\documentclass[draft]{agujournal2019}
\usepackage{url} 
\usepackage{lineno}
\usepackage[inline]{trackchanges} 
\usepackage{soul}

\draftfalse

\journalname{JGR: Space Physics}

\begin{document}

\title{The Io, Europa and Ganymede auroral footprints at Jupiter in the ultraviolet: positions and equatorial lead angles}

%
%


\authors{V. Hue\affil{1,2}\thanks{vincent.hue@lam.fr},
G. R. Gladstone\affil{2,3},
C. K. Louis\affil{4},
T. K. Greathouse\affil{2},
B. Bonfond\affil{5},
J. R. Szalay\affil{6},
A. Moirano\affil{7,8},
R. S. Giles\affil{2},
J. A. Kammer\affil{2},
M. Imai\affil{9},
A. Mura\affil{7},
M. H. Versteeg\affil{2},
G. Clark\affil{10},
J.-C. G\'erard\affil{5},
D. C. Grodent\affil{5},
J. Rabia\affil{11},
A. H. Sulaiman\affil{12},
S. J. Bolton\affil{2},
J.E.P. Connerney\affil{13,14}}

\affiliation{1}{Aix-Marseille Université, CNRS, CNES, Institut Origines, LAM, Marseille, France}
\affiliation{2}{Southwest Research Institute, San Antonio, Texas, USA}
\affiliation{3}{University of Texas at San Antonio, San Antonio, Texas, USA}
\affiliation{4}{School of Cosmic Physics, DIAS Dunsink Observatory, Dublin Institute for Advanced Studies, Dublin, Ireland}
\affiliation{5}{STAR Institute, LPAP, Universit\'e de Li\`ege, Li\`ege, Belgium}
\affiliation{6}{Department of Astrophysical Sciences, Princeton University, Princeton, NJ, USA}
\affiliation{7}{Institute for Space Astrophysics and Planetology, National Institute for Astrophysics, Rome, Italy}
\affiliation{8}{Sapienza University of Rome, Rome, Italy}
\affiliation{9}{Department of Electrical Engineering and Information Science, National Institute of Technology (KOSEN), Niihama College, Niihama, Japan}
\affiliation{10}{Johns Hopkins University Applied Physics Laboratory, Laurel, MD, USA}
\affiliation{11}{Institut de Recherche en Astrophysique et Planétologie, CNRS-UPS-CNES, Toulouse, France}
\affiliation{12}{Minnesota Institute for Astrophysics, School of Physics and Astronomy, University of Minnesota, Minneapolis, MN, United States}
\affiliation{13}{NASA Goddard Spaceflight Center, Greenbelt, MD 20771, USA}
\affiliation{14}{Space Research Corporation, Annapolis, MD 21403, USA}

\correspondingauthor{V. Hue}{vincent.hue@lam.fr}

\begin{keypoints}
\item Over 1600 ultraviolet spectral images of the Io, Europa and Ganymede footprints from Juno are analyzed
\item Empirical formulae for the Io, Europa and Ganymede lead angles derived from Juno data are provided
\item Alfv\'en travel time estimates are derived, constraining the Alfv\'enic interaction at the 3 innermost Galilean moons
\end{keypoints}

%
%

%
%


\begin{abstract}
Jupiter’s satellite auroral footprints are a consequence of the interaction between the Jovian magnetic field with co-rotating iogenic plasma and the Galilean moons. The disturbances created near the moons propagate as Alfv\'en waves along the magnetic field lines. The position of the moons is therefore "Alfv\'enically" connected to their respective auroral footprint. The angular separation from the instantaneous magnetic footprint can be estimated by the so-called lead angle. That lead angle varies periodically as a function of orbital longitude, since the time for the Alfv\'en waves to reach the Jovian ionosphere varies accordingly. Using spectral images of the Main Alfv\'en Wing auroral spots collected by Juno-UVS during the first forty-three orbits, this work provides the first empirical model of the Io, Europa and Ganymede equatorial lead angles for the northern and southern hemispheres. Alfv\'en travel times between the three innermost Galilean moons to Jupiter's northern and southern hemispheres are estimated from the lead angle measurements. We also demonstrate the accuracy of the mapping from the Juno magnetic field reference model (JRM33) at the completion of the prime mission for M-shells extending to at least 15\,R$_J$. Finally, we shows how the added knowledge of the lead angle can improve the interpretation of the moon-induced decametric emissions.
\end{abstract}

\section*{Plain Language Summary}
The interaction between the Jovian magnetospheric plasma and the Galilean moons gives rise to a complex set of phenomena, including the generation of auroral spots magnetically related to the moons and the generation of radio emissions. The magnetic perturbations local to the moons propagate at a finite speed along the magnetic field lines, and reach the northern and southern Jovian hemispheres where they produce the auroral spots. Studying the position of these auroral spots and how they vary over a complete Jovian rotation provides information about the magnetic mapping, as they map directly to the actual physical positions of the moons. The magnetic field model derived from Juno's prime mission is in good agreement with the observation of the satellite footprints. This paper provides information about how the electromagnetic perturbation resulting from the interaction propagates at a finite speed to create auroral spots, leading to an angular shift between the instantaneously magnetically-mapped position of the moon and the auroral footprint, a quantity also known as the "equatorial lead angle". The present work provides an empirical fit of the equatorial lead angle for Io, Europa and Ganymede derived from Juno data.

\section{Introduction}

Jupiter's volcanically-active moon Io is a major source of plasma within the Jovian magnetosphere. Over a ton per second of sulfur dioxide escapes from the moon as a neutral cloud around the moon. High-energy electrons trapped in Jupiter's magnetosphere dissociate and ionize the neutral material, producing ions which get picked up by Jupiter's strong rotating magnetic field, forming the Io plasma torus. The plasma is then radially transported outwards over the timescale of several weeks as a plasma sheet with a typical scale height of about 1\,R$_J$ (1 R$_J$ = 71492\,km), and confined around the centrifugal equator, near the magnetic equator, \textit{i.e.,} the points along the magnetic field lines that are farthest from the spin axis of Jupiter \cite <e.g.,>[]{Acuna1983, Belcher1983}.

Evidence of the magnetospheric interaction at Io was discovered early on through the detection of decametric radio emission \cite{Bigg1964}. The subsequent Voyager 1 measurements near Io led to formalize the interaction model at the moon resulting in the generation of an Alfv\'enic disturbance \cite{Neubauer1980, Goldreich1969, Acuna1981, Belcher1981}. These Alfv\'en waves propagate to Jupiter along the magnetic field lines towards both polar regions, and their multiple reflections off of the plasma density gradients along the way were later proposed to be responsible for the multiple Io-controlled decametric radio arcs observed \cite{Gurnett1981}.

When the Alfv\'enic perturbations reach high Jovian latitudes, they accelerate electrons towards and away from Jupiter, as well as protons, leading to the generation of the auroral footprints and decametric radio waves \cite{Hess2010, Szalay2020c}, which were initially detected in the infrared (IR) and later in the ultraviolet (UV) \cite{Connerney1993, Clarke1996, Prange1996}. Detection of the Europa and Ganymede auroral footprints using the Hubble Space Telescope (HST) followed shortly after \cite{Clarke2002}.

The overall strength of the moon-magnetosphere interaction depends on the magnetic field strength, the plasma density, the conductivity, as well as the electron acceleration efficiency \cite{Saur2013, Sulaiman2023}. Because Jupiter's magnetic dipole is tilted from its spin axis by about 10 degrees, the plasma sheet sweeps through the Galilean moons twice every $\sim$10\,hour-rotation period. In the reference frame centered about the plasma sheet, the moons travel up and down the sheet, experiencing denser plasma condition at its the center. The multiple reflections of the Alfv\'en waves on Alfv\'en speed gradients, such as near the ionosphere, create multiple auroral spots, whose relative position, morphology, and distribution is modulated by the moon centrifugal latitude, as reported by remote sensing observations \cite <e.g.,>[]{Bonfond2008, Mura2018}.

Extensive campaigns using the Hubble Space Telescope (HST) have allowed categorizing the variable morphology of the Io footprints structure as a function of the location of Io with respect to the plasma sheet \cite{Gerard2006, Bonfond2008, Wannawichian2013}. These studies showed that the Main Alfv\'en Wing (MAW) spot was sometimes preceded by a leading spot, named the Transhemispheric Electron Beam (TEB) spot, for a very specific range of sub-Io Jovian longitudes. In that case, the Alfv\'en waves generated in the wing of the interaction region later reached Jupiter's ionosphere. This means the beam of electrons generating the TEB spot was accelerated anti-planetward by the MAW on the opposite hemisphere, then travelled along the magnetic field line throughout the torus, unaffected by the higher plasma torus density there \cite{Bonfond2008, Hess2013}.

The Juno mission has brought a wealth of new observations that significantly enhanced our understanding of the moon-magnetosphere interaction \cite{Bolton2017, Connerney2017}. Juno crosses the magnetic shells connected to the orbits of Io, Europa and Ganymede at least twice per orbit. Not only did Juno measure the field and particles within the magnetic fluxtube connected to the discrete auroral footprint spots or tail, but it also provided unprecedented infrared and ultraviolet observations both at the highest spatial resolution ever achieved \cite{Mura2018}, and with viewing geometries not accessible from HST \cite{Hue2019b}. 

Prior to Juno, there was a debate regarding the processes responsible for the footprint tail emission. One set of studies favored a quasi-steady current system transferring angular momentum from the Jovian ionosphere to the sub-corotating plasma in the moon wakes, and which would be characterized by beams of electron discrete in energies \cite <e.g.,>[]{Goldreich1969, Hill2002, Delamere2003, Su2003}. Other studies suggested it results from the multiple bouncing of the Alfv\'en waves generated at the moons, which would be characterized by bidirectional electron populations with broadband energy distributions \cite <e.g.,>[]{Bonfond2009, Hess2010, Bonfond2017b, Crary1997, Jacobsen2007}.

Early Juno measurements demonstrated the Alfv\'enic nature of the interaction associated with the Io footprint tail \cite{Szalay2018, Damiano2019, Gershman2019, Sulaiman2020, Sulaiman2023, Szalay2020b}. \citeA{Szalay2018, Szalay2020b} showed that the electron energy flux, obtained by the multiple Juno-JADE measurements along the Io footprint tail, is a function of the angular separation along Io's orbit between the moon and the MAW spot tracked from Jupiter's ionosphere to the satellite orbit. This angle is also known as "Io-Alfv\'en tail distance". Juno measured upward ion conics, \textit{i.e.} protons with angular distribution concentrated along the loss-cone, detected simultaneously with ion cyclotron waves, showing that energetic ions are also generated from moon-magnetosphere interaction \cite{Clark2020}, as well as low-energy ion acceleration in the vicinity of the Main Alfv\'en Wing, both near the ionosphere and near the Io torus boundary \cite{Szalay2020c}. Additionally, the cyclotron maser instability (CMI) driven by a loss-cone distribution has been established as a major process at Jupiter in generating hectometric and decametric emissions, induced or not by the Galilean moons \cite{Louarn2017, Louarn2018, Louis2020}.

Alfv\'enic acceleration processes were also observed when crossing the fluxtube connected to Ganymede's footprint tail \cite{Szalay2020a}. One particular Ganymede fluxtube crossing (on PJ30, 8 Nov. 2020), during which Juno was connected for the first time to the leading-most Ganymede auroral spot, brought a set of in-situ and remote sensing measurements consistent with what was expected during a TEB crossing \cite{Hue2022}. Unlike for Io and Ganymede, crossing through the fluxtube connected to the Europa footprint tail showed signs of electron distribution resulting at least in part from electrostatic acceleration processes \cite{Allegrini2020}. Whether this is a fundamental difference of the interaction at Europa, or that it actually also corresponded to a TEB crossing remains to be ascertained by additional studies of the Europa footprint tail crossings. Radio emissions are also observed associated with Ganymede's interaction \cite{Louis2020}.

The scope of this paper is to process the ultraviolet auroral footprint observations performed from the first perijove (PJ, hereafter) on 8 Aug 2016 until PJ43 (5 Jul 2022) with the ultraviolet spectrograph on Juno (Juno-UVS). Section \ref{sec:JunoUVS} describes the observations and data reduction procedure. In section \ref{sec:sat_fp}, the reported satellite footprint locations are compared against the predicted satellite footpaths from the magnetic field model JRM33 \cite{Connerney2022}. From the position of the MAW spot of the satellites, the equatorial lead angle can be estimated using a magnetic field model. The equatorial lead angle is the angular shift, calculated along the orbital motion of the moons, between the actual position of the moons and the position of the MAW footprint mapped instantaneously to the orbital plane of the moon. We then calculate the equatorial lead angles for Io, Europa and Ganymede and provide an empirical fit on section \ref{sec:LA}, as well as estimates of the Alfv\'en travel times for each of the three moons. We subsequently discuss the lead angle variation measured in section \ref{sec:variability}. Finally, in section \ref{sec:decametric}, we present an additional example on the use of the lead angle for the interpretation of the Ganymede-induced decametric emission.

\section{Juno-UVS observations}
\label{sec:JunoUVS}

Juno is a spin-stabilized spacecraft placed in a polar and highly eccentric orbit around Jupiter since July 2016 \cite{Bolton2017}. Each orbit, Juno performs a close flyby of the northern polar region first, reaches closest approach at lower Jovian latitudes at an altitude of about 4000\,km, and then flies above the southern polar region. Because Juno's orbit is highly elliptical and because its closest approach velocity with respect to Jupiter is about 58\,km/s, it takes Juno about 2 hours to fly from north to south pole. This implies that the spatial resolution of the remote sensing instrument such as UVS varies drastically over the course of a perijove observation sequence. As the mission is progressing, the orbital period, which was initially around 53 days, was reduced to a shorter orbit, and will be reduced down to 33 days around PJ75 (15 Aug 2025). Each major orbital period reduction follows a Galilean moon flyby. Juno's orbit precesses over time and the sub-spacecraft PJ latitude increases as the mission continues. The viewing window of UVS over the northern and southern auroras grows increasingly more asymmetric with the mission, with the viewing window of the northern aurora decreasing over time.

Some of the magnetospheric goals of Juno include performing in-situ measurements of the particle population in Jupiter's magnetosphere using an electron and ion sensors suite, while remotely sensing the associated infrared and ultraviolet aurora they may trigger on Jupiter or on the Galilean moons \cite{Bagenal2017}. The Ultraviolet Spectrograph (UVS) is a photon-counting imaging spectrograph operating in the 68-210\,nm range \cite{Gladstone2017_SSR}. Each spin of Juno, UVS records a swath of UV emission along its 7.2$^{\circ}$-long slit, with a typical point-source integration time of 17\,ms. The point-spread function (PSF) and spectral resolution are respectively 0.1$^{\circ}$ and 1.3\,nm, at best \cite{Davis2011, Greathouse2013}, meaning that UVS can resolve features on Jupiter down to 60\,km at PJ. Because Juno is flying above both auroral regions at higher altitude, UVS can rather resolve features in the $\sim$150-400\,km range there. Counts recorded on the detector are then converted into brightness using the instrument effective area derived from thousands of stellar observations recorded throughout the mission in between PJ observation sequences \cite{Hue2019a, Hue2021b}.

UVS is equipped with a scan mirror that allows its field of regard to be shifted up to $\pm$30$^{\circ}$ away from the spin plane. Juno's spinning nature combined with UVS' mirror pointing capability allows building up complete maps of Jupiter's aurora by co-adding consecutive swaths of data. The best temporal resolution of UVS is constrained by the spin rate of Juno and cannot be less than 30\,seconds. 

The satellite auroral footprints represent a direct magnetic mapping to a given orbital distance, through the prism of the sub-Alfv\'enic interaction around each moon. It is worth differentiating at this point magnetic mapping from Alfv\'enic mapping. The Main Alfv\'en Wing spot is Alfv\'enically connected to the physical position of the moon, which differs from the instantaneous field-line tracing magnetically connecting the moon and Jupiter's ionosphere. Because of the Jovian rotation and the moon orbital motion, there is a displacement of the satellite footprints location between two consecutive UVS images recorded 30\,seconds apart, which can be estimated considering each moon's synodic period:

\begin{equation}
 P^{syn}_{moon} = \frac{P_{J} \, P_{moon}}{P_{moon} - P_{J}},   
 \label{eq:eq1}
\end{equation}
 
where P$_{J}$ and P$_{moon}$ are respectively the rotation period of Jupiter and orbital periods of a moon. Relationship (\ref{eq:eq1}) leads to P$^{syn}_{io}$ = 12.89\,h; P$^{syn}_{europa}$ = 11.22\,h; P$^{syn}_{ganymede}$ = 10.53\,h. Over the course of a Juno spin ($\sim$30\,seconds), the moon spans a longitudinal sector of $\Delta \lambda_{moon} = (360 / P^{syn}_{moon} ) \times 30$, where $P^{syn}_{moon}$ is in seconds. This leads to $\Delta \lambda_{io}\,=\,0.23^{\circ}$; $\Delta \lambda_{europa}\,=\,0.27^{\circ}$; $\Delta \lambda_{ganymede}\,=\,0.28^{\circ}$ between two consecutive swaths. Over one Juno spin and using the JRM33 model, this leads to a displacement of the instantaneous field line from the moon to Jupiter's northern ionosphere up to 223\,km, 220\,km, 191\,km, respectively for Io, Europa and Ganymede. In the south, that same smearing ranges up to 141\,km, 136\,km, 114\,km for Io, Europa and Ganymede. Note that all longitudes ($\lambda$) quoted in this work are west longitudes in the System III reference frame, and are calculated according to the 1965 system 3 rotation period \cite{Dessler1983}.

\begin{figure}[h!]
\centering
\includegraphics[width=0.95\columnwidth]{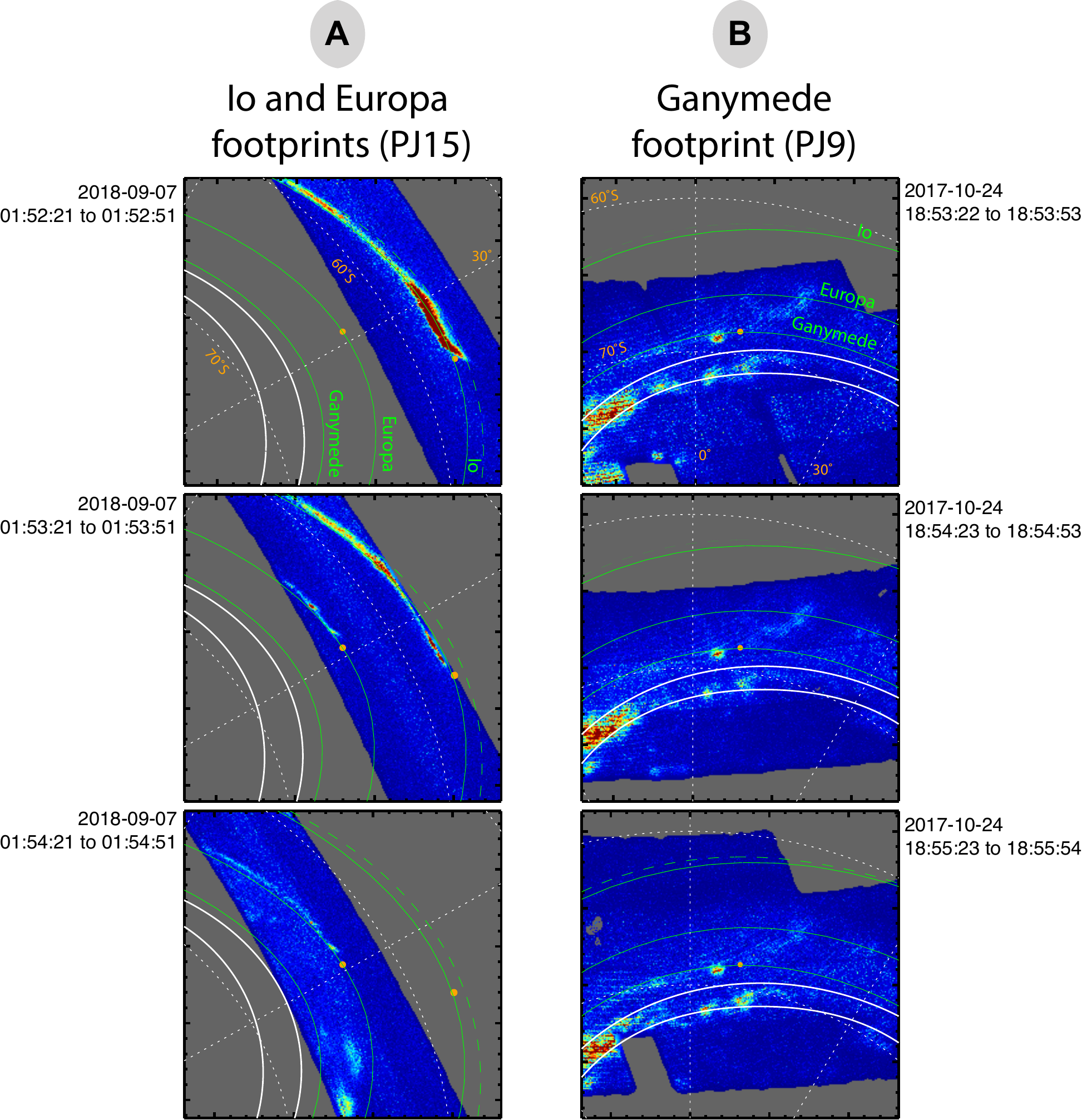}
\caption{2-spin averaged Juno-UVS spectral images of the Io and Europa footprints (panel A), and the Ganymede footprints (panel B) over the southern hemisphere. The UVS nadir time of the two consecutive spins are listed as UTC time. The instantaneous moon magnetic footprint positions according to the JRM33 model is shown as orange dots along the satellite footprints, also calculated using JRM33 and shown as solid green line. The satellite footpath of Io from \citeA{Bonfond2017} is shown as dashed green lines. The solid white lines show the reference location where the main oval is observed \cite{Bonfond2012}.}
\label{fig:Fig_UVS}
\end{figure}

In order to increase the signal to noise ratio (SNR), without smearing the signal of the satellite footprint emission, the choice was made to bin the UVS data by 2 consecutive spins worth of data. Various auroral emissions may be present close to the footprints, such as emission from the main oval \cite{Bonfond2021, Ebert2021}, injection signatures \cite <e.g.,>[]{Bonfond2017}, or increased background radiation \cite{Bonfond2018, Kammer2019}. The change of geometry, background emission and SNR makes challenging extracting the precise location of the footprints, particularly for the Europa and Ganymede footprints, generally located rather close to the main oval emission and auroral injections. For these reasons, the footprint locations were manually extracted by visual inspection of consecutive series of 2 spins-averaged UVS data. Since the moons are not in co-rotation with Jupiter, identifying the satellite footprints over consecutive spins worth of data is achieved by visually inspecting auroral spots that are in sub-corotation around a background of mostly corotating auroral emission. Figure \ref{fig:Fig_UVS} shows an example of consecutive 2-spin averaged spectral image recorded by UVS of the Io, Europa and Ganymede footprints.

\begin{figure}[h!]
\centering
\includegraphics[width=0.99\columnwidth]{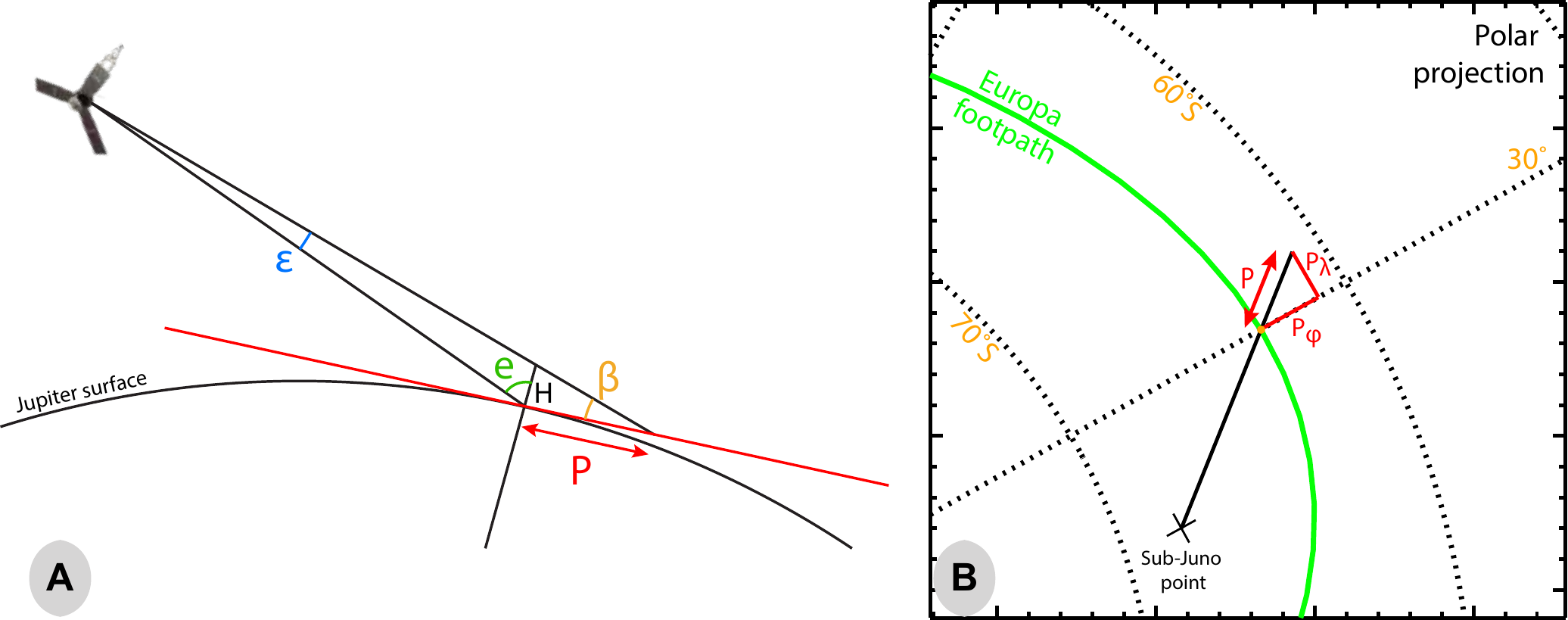}
\caption{Uncertainty calculation on the derived latitude and longitude of the footprints. Panel A: calculation of the projected height of a typical auroral footprint vertical emission P along the Juno line of sight. The latitude and longitude uncertainty ($P_\phi$, $P_\lambda$) are calculated through the decomposition of $P$, exaggerated here for illustration purposes, along the latitude and longitude grid (panel B). $e$ is the emission angle, while $\beta$ and $\epsilon$ are two angles used to calculate the projection of $P$.}
\label{fig:geometry}
\end{figure}

The uncertainties in the derived location of the different observed footprints were calculated as the quadratic combination of the uncertainty due to the instrument point-spread-function (PSF), and the uncertainty due to the projected scale height $H$ of the footprint emission curtain. While the former uncertainty only depends on the distance between Juno and the footprint of interest, the latter depends on the footprint emission angle as seen by Juno. The uncertainties on the footprint longitude ($\lambda$), latitude ($\phi$) therefore reads: 

\begin{equation}
 \sigma_{\lambda, \phi}^2 = \sigma_{PSF}^2 + \sigma_{P \,\lambda\,\phi} ^2,  
 \label{eq:2}
\end{equation}

where $\sigma_{PSF}$ is calculated as the projection of the $\sim$\,0.1$^{\circ}$ UVS PSF along the latitude and longitude grid. $\sigma_{P \,\lambda\,\phi}$ is calculated as the projected extent of the footprint emission curtain, $H$, as illustrated on figure \ref{fig:geometry}, assuming as a first approximation that the auroral curtain is vertical. P is given as P = H/tan\,$\beta$, and $\beta$ = $\pi/2$ - $e$ - $\epsilon$, where $e$ (emission angle) and $\epsilon$ are angles calculated using NAIF's SPICE kernels \cite{Acton1996, Acton2018}. $H$ was taken as 366\,km, \textit{i.e.,} the typical scale height of a Chapman profile derived from Io's MAW spot by \citeA{Bonfond2010}.

Only the data recorded by UVS when Juno was 1.5\,hours about perijove was used, \textit{i.e.,} when the $\sim$\,0.1$^{\circ}$ PSF projected at 45$^{\circ}$ on Jupiter's surface was lower than the typical width of the footprints as previously derived from HST \cite{Bonfond2010}. From PJ1 until PJ43, this results in a set of 211 (Io), 108 (Europa) and 160 (Ganymede) 2-spin averaged UVS images in the north, and 585, 264 and 299 in the south for Io, Europa and Ganymede, respectively. The projection altitude of the UV data on Jupiter used for this work was 900\,km above the 1-bar level, after \citeA{Bonfond2010, Szalay2018}.

\section{Satellite footprint locations}
\label{sec:sat_fp}

Measuring accurately the location of the satellite footprint emission gives a precise magnetic mapping of the orbital position of the moons. Comparing the observed footprint position of, \textit{e.g.,} Io, against the predicted magnetic mapping of that moon gives an estimate of the mapping accuracy. Earlier magnetic field models of Jupiter were constructed using in-situ spacecraft measurements as well as the Io footprint location derived from IR and UV observations. This initially allowed \citeA{Connerney1998} to constrain the VIP4 model, using observations of the Io footprints in the IR. When employing a larger set of Io footprint observations obtained in the UV, the method was extended by \citeA{Hess2011} to produce the VIPAL magnetic field model. That model additionally included the longitudinal constraint on the Io footprint location, which originates from the finite Alfv\'en travel time between the interaction region near the moon to Jupiter's ionosphere. Later, \citeA{Hess2017} further constrained Jupiter's magnetic field model by adding also Europa's and Ganymede's auroral footprint locations to produce the ISaAC magnetic field model.

\begin{figure}[h!]
\centering
\includegraphics[width=1.1\columnwidth]{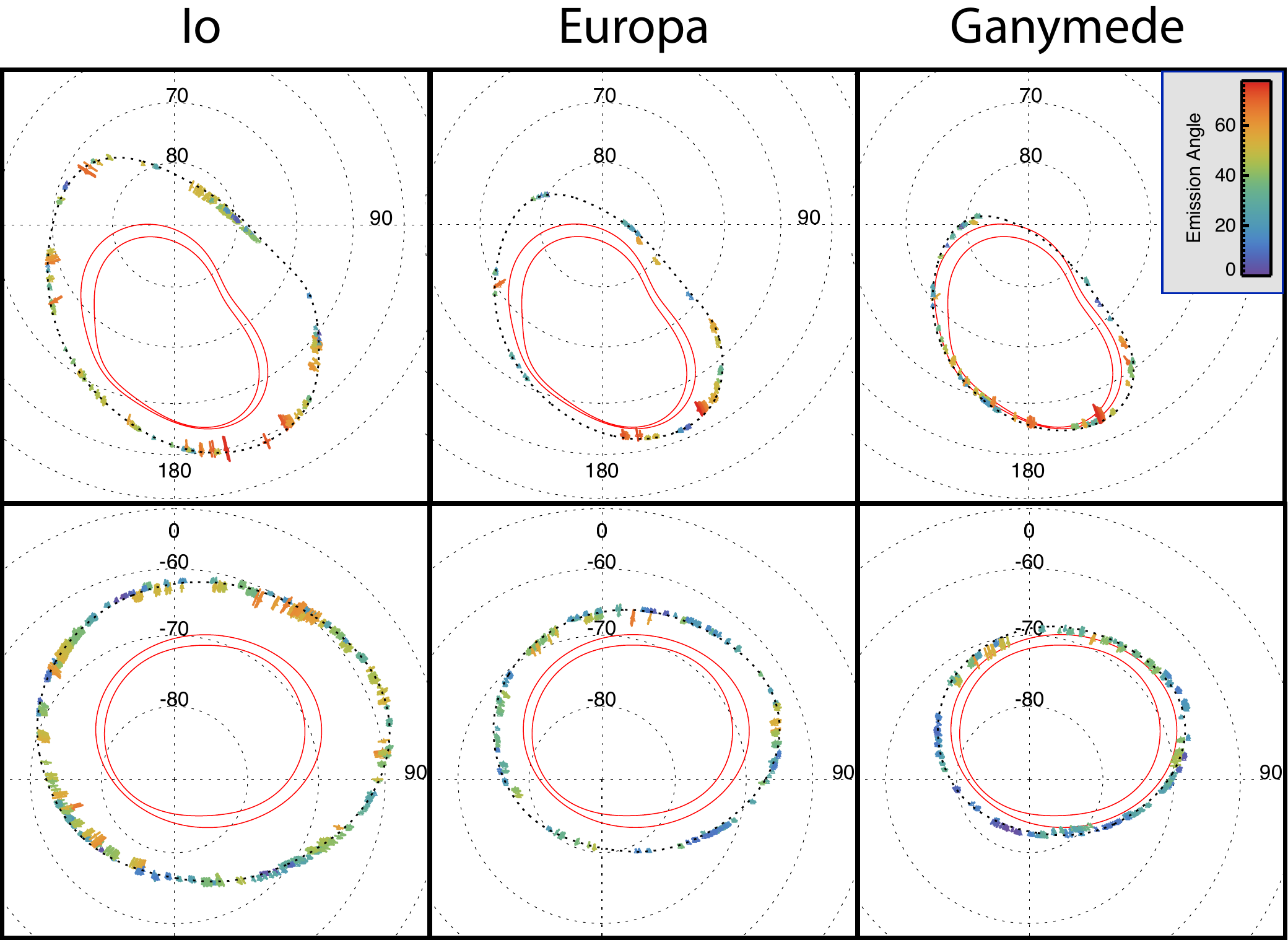}
\caption{Extracted position of the Io, Europa and Ganymede Main Alfv\'en Wing footprints in the north (top panels) and south (bottom panels), using Juno-UVS data recorded from perijoves 1 to 43 and color-coded according to the emission angle observed by Juno-UVS. The solid red lines show the reference oval from \citeA{Bonfond2012}. The black dotted line show the satellite footpaths as predicted from the JRM33 model \cite{Connerney2022}.}
\label{fig:FP_locations}
\end{figure}

The position of the Io, Europa and Ganymede footprints are presented in figure \ref{fig:FP_locations}, and are plotted against the satellite footpath contours predicted from the JRM33 model \cite{Connerney2022}, combined with the Juno-era current sheet model \cite{Connerney2020}. Unlike the HST observations, UVS observed the footprints from a wide range of emission angles ($e$ on figure \ref{fig:geometry}) from 1$^{\circ}$ to 77$^{\circ}$, as well as from a wide range of altitude, from 0.3\,R$_J$ to 2.6\,R$_J$.

Because of the precession of Juno's orbit, Juno-UVS gets an increasingly shorter look over the northern aurora as the mission progresses. This causes a greater density of footprint position measurements in the south, despite the rapid increase of Juno's altitude over the south pole during the outbound leg. The magnetic field model obtained after the completion of the prime mission (JRM33) predicts a magnetic footpath for the satellite in very good agreement with the Juno-UVS observations \cite{Connerney2022}. One region which previously showed significant differences between the observed and predicted satellite footprint location is the auroral kink sector, at longitudes from 80$^{\circ}$ to 150$^{\circ}$ in the northern hemisphere only, for which \citeA{Grodent2008} suggested the inclusion of a magnetic anomaly to the existing magnetic field model at that time, in order to better fit the footprint location. Early Juno measurements for instance showed significant differences between the predicted Europa footprint location in that sector, and the observed one \cite{Allegrini2020b}. JRM33 is now in very good agreement with the observed MAW spot positions in that sector for all 3 inner Galilean moons.

Despite the more extended UVS coverage than HST in the south, there are several longitude gaps in the northern MAW observation not covered by Juno-UVS as of PJ43. For the northern Io MAW spot, there are gaps at footprint longitudinal sectors of 260$^{\circ}$-280$^{\circ}$; 100$^{\circ}$-118$^{\circ}$; 342$^{\circ}$-20$^{\circ}$. For the northern Europa MAW, the gaps are 300$^{\circ}$-80$^{\circ}$; 175$^{\circ}$-206$^{\circ}$; 215$^{\circ}$-238$^{\circ}$; 250$^{\circ}$-290$^{\circ}$. For the northern Ganymede MAW, the main longitude gap is 280$^{\circ}$-130$^{\circ}$. The MAW footprint latitude/longitude locations for Io, Europa and Ganymede are provided in the supporting information. The difference in coverage with previous HST observations can be compared with figures 1 and 2 showed in the supporting information of \citeA{Bonfond2017b}.

By binning the measured MAW positions over 1.5$^{\circ}$-wide longitudinal sector to minimize the measurements scattering, and only considering the MAW spot position recorded at emission angles lower than 20$^{\circ}$, one can assess the accuracy of magnetic field models, such as JRM33, while limiting the uncertainty resulting from the range of viewing geometries. For Io, the average distance between the JRM33-computed reference footpath and the observed MAW positions is 511\,$\pm$\,28\,km in the north and 274\,$\pm$\,64\,km in the south. For Europa, these numbers are 141\,$\pm$\,26\,km in the north and 322\,$\pm$\,61\,km in the south. For Ganymede, these numbers are 213\,$\pm$\,44\,km in the north and 343\,$\pm$\,64\,km in the south. These numbers provide an order-of-magnitude estimate in mapping uncertainty between the model and the observations, and vary slightly ($\sim$10\%) depending on the way the UVS data is binned (\textit{e.g.,} longitudinal and emission angle binning).

The position of the MAW spot of the moons, together with a magnetic field model, is a key ingredient to estimate the equatorial lead angle, which is the focus of the next section.

\section{Lead angles}
\label{sec:LA}

The equatorial lead angle ($\delta$) is the angular separation between the position of the moon and the magnetically-mapped MAW footprint onto the orbital plane. The lead angle is determined by the sum of the physical processes occurring between the interaction region around the moon and Jupiter's ionosphere. In this work that uses Juno-UVS data, it is calculated by:

\begin{enumerate}
    \item measuring the position of the MAW spot at a given time t$_0$,
    \item determining the position of the moons at the same time t$_0$ from the ephemerides,
    \item tracing back the position of the MAW spot in the moon orbital plane using a magnetic field model,
    \item calculating the longitudinal difference between the moon and the back-traced MAW spot, hence the equatorial lead angle.
\end{enumerate}

Juno in-situ instruments provided invaluable measurements of the particle distribution on the magnetic field lines connected to the satellite footprints and tails, such as measurements connected to Io \cite{Szalay2018, Clark2020, Sulaiman2020, Sulaiman2023, Szalay2020b}, Europa \cite{Allegrini2020} and Ganymede \cite{Szalay2020a, Louis2020, Hue2022}. \citeA{Szalay2020b} showed that the exponential decrease in precipitating electron energy flux, obtained from the Juno-JADE instrument when Juno was magnetically connected at various distances downstream the Io footprint tail, was better organized when considering the angular separation along Io's orbit between Io and an Alfv\'en wave back-traced from Jupiter's ionosphere to the moon orbital plane, as the power decays farther from the moons for each subsequent bounce of the initial Alfv\'en wave generated near the moon. Further, \citeA{Sulaiman2023} showed the Poynting fluxes and field-aligned currents to similarly exhibit a decay as a function of the same downtail angle. That quantity, coined "Io-Alfv\'en tail distance" by \citeA{Szalay2020b}, wraps up the knowledge of the equatorial lead angle within. Here, we provide critical updates unique to Juno's observational platform allowing for the calculation of Alfv\'en angles for Io, Europa, and Ganymede.

\begin{figure}[h!]
\centering
\includegraphics[width=0.8\columnwidth]{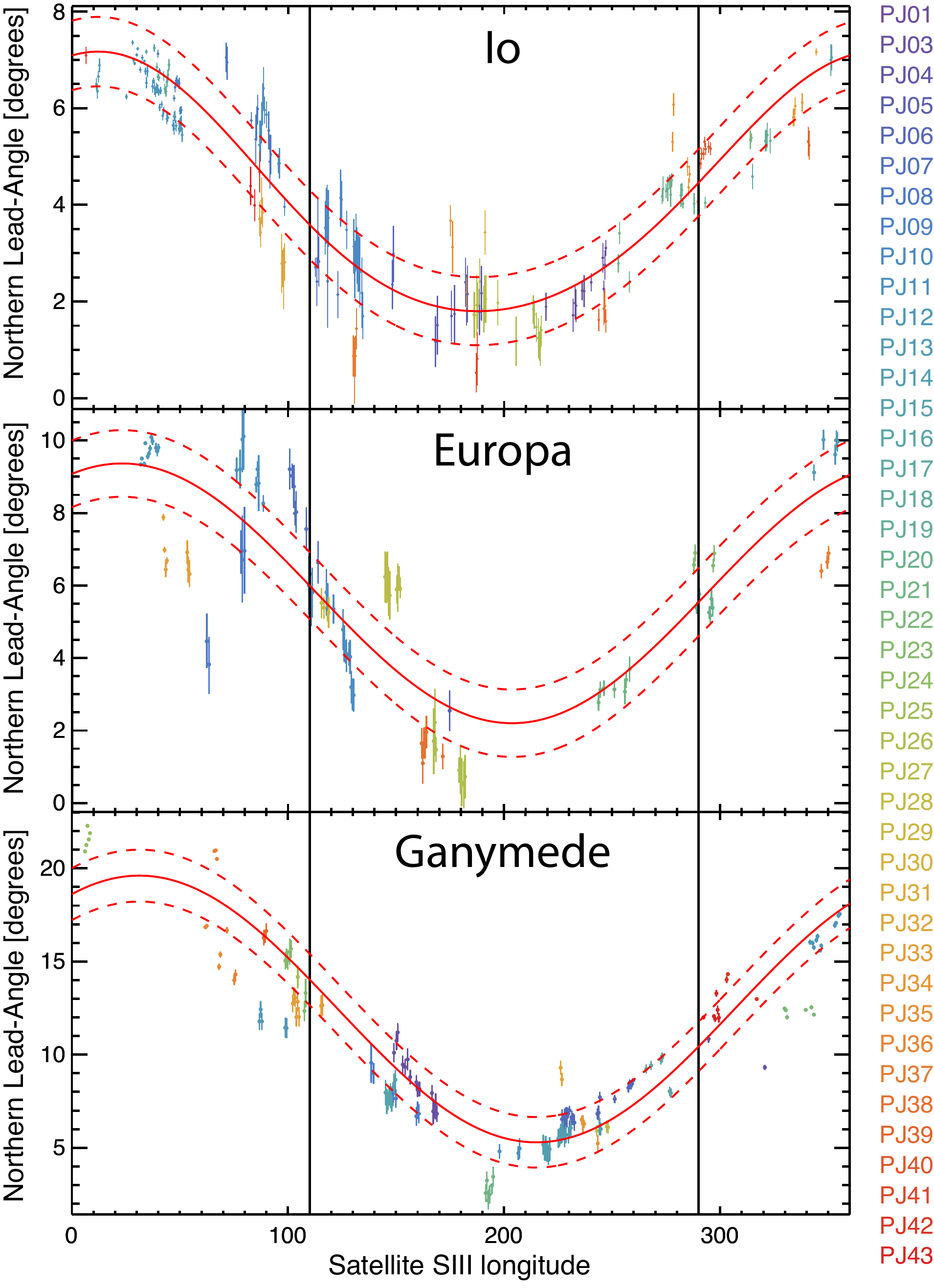}
\caption{Northern lead angle for Io (top), Europa (middle) and Ganymede (bottom) using Juno-UVS dataset of the Main Alfv\'en Wing from PJ1 until PJ43. The vertical solid black lines indicate the location when the moons are at the center of the plasma sheet. The red lines correspond to the best fit obtained with eq. (\ref{eq:fourier}). The red-dashed lines correspond to the 3$\sigma$ uncertainty on the fit parameters.}
\label{fig:Northern_LA}
\end{figure}

\begin{figure}[h!]
\centering
\includegraphics[width=0.8\columnwidth]{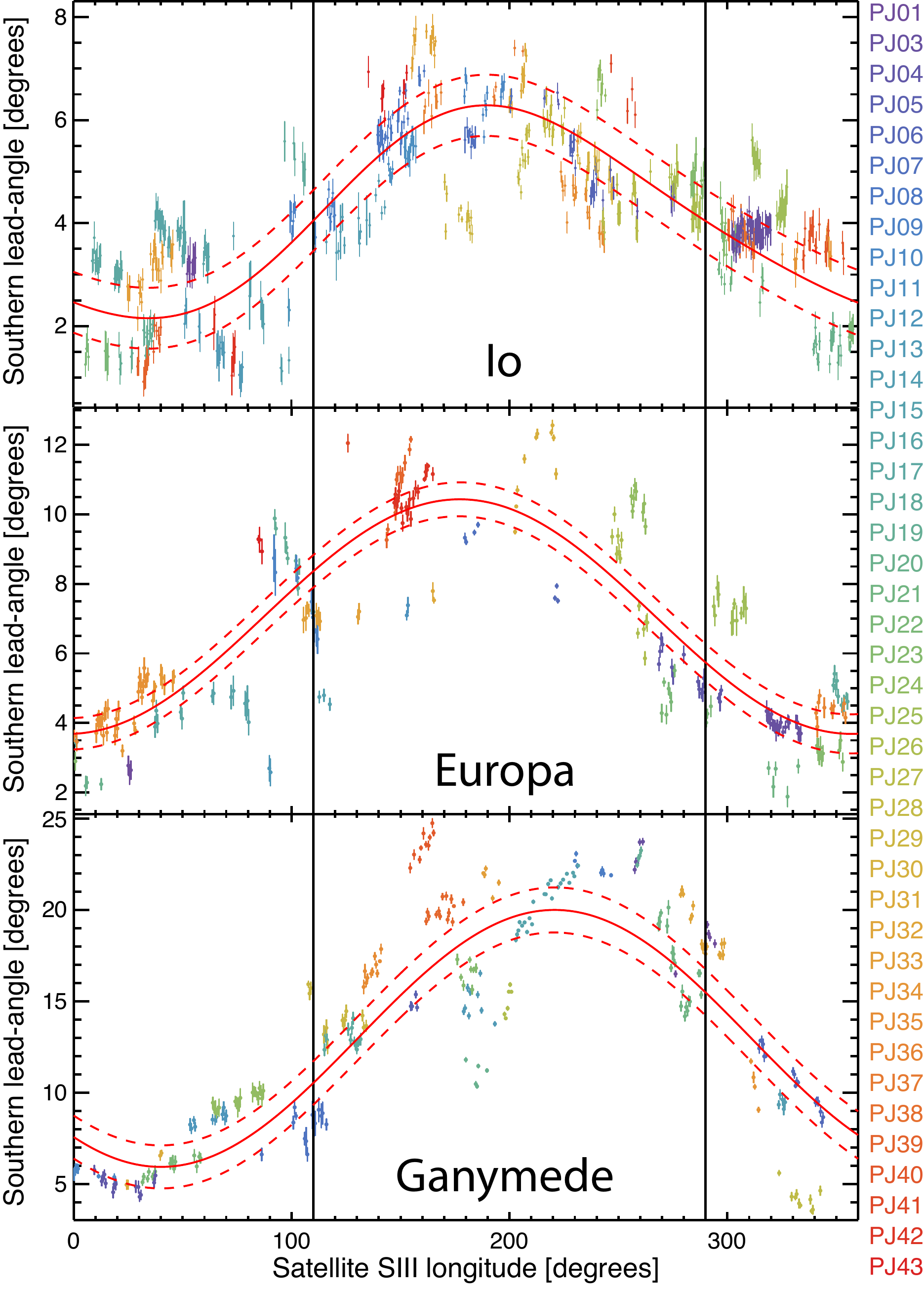}
\caption{Same as figure \ref{fig:Northern_LA} for the south.}
\label{fig:Southern_LA}
\end{figure}

Figures \ref{fig:Northern_LA} and \ref{fig:Southern_LA} display the equatorial lead angle in the north and south, respectively, inferred from Juno-UVS and using the JRM33 magnetic field model \cite{Connerney2022} combined with the Juno-era current sheet model \cite{Connerney2020}, and color-coded according to the perijove it was recorded at. The lead angle depends mainly on the travel time variation of the Alfv\'en waves through the higher plasma density region of the plasma sheet. Assuming a plasma sheet around Jupiter with uniform and constant plasma density to first order, one would expect a sinusoidal modulation of that angle with the latitudinal separation between the moons and the central part of the plasma sheet over the course of a Jovian rotation \cite{Hess2010b}. Figures \ref{fig:Northern_LA} and \ref{fig:Southern_LA} are adjusted with a Fourier series, such that:

\begin{equation}
  \delta = a_0 + \sum_{n=1}^{\infty} a_n\,\cos n w \lambda + b_n\,\sin n w \lambda \label{eq:fourier}
\end{equation}

The best fit parameters of equation \ref{eq:fourier} were estimates using a least-squares minimization technique \cite <MPFIT package,>[]{Markwardt2009}, and are listed on Table \ref{tab:fit_param}. In order to assume periodicity in the fit, the measured lead angles were replicated in the \mbox{$[-2\pi:4\pi]$} interval. Because of the sparsity of the Europa and Ganymede data, fitting the data with higher order Fourier series introduces oscillations. The lead angle data on these two moons is therefore fitted using a first order Fourier series. For Io, a second order Fourier series best reproduces the observations, based on the weighted sum of squared residuals, though producing larger uncertainties of the retrieved fit parameters.

The averaged values of the northern and southern Io lead angles are 4.3$\,\pm 0.7$\,degrees and 4.1$\,\pm 0.6$\,degrees, respectively. For Europa, these values are 5.8$\,\pm 0.9$\,degrees in the north, and 7.0$\,\pm 0.5$\,degrees in the south. For Ganymede, the averaged northern and southern lead angles are 12.3$\,\pm 1.3$\,degrees and 13.0$\,\pm 1.3$\,degrees, respectively. The uncertainties in the averaged lead angle is calculated using the 3-sigma uncertainties of the derived fit parameters, and are tighter in the south, due to the denser dataset. The averaged northern and southern lead angles value overlap for each individual moon, meaning that the Alfv\'en travel time is similar between hemispheres.

\begin{table}[ht]
\begin{center}       
\begin{tabular}{c|c|c} 
 & North & South  \\
 \hline
Io & $\delta$ = 4.26 + 2.64 $\cos$ $\lambda$ + 0.50 $\sin$ $\lambda$  & 
$\delta$ = 4.14 - 1.89 $\cos$ $\lambda$ - 0.70 $\sin$ $\lambda$  \\
 & + 0.20 $\cos$ 2\,$\lambda$ + 0.126 $\sin$ 2\,$\lambda$ & + 0.22 $\cos$ 2\,$\lambda$ - 0.12 $\sin$ 2\,$\lambda$  \\
Europa & $\delta$ = 5.78 + 3.29 $\cos$ 0.99 $\lambda$ + 1.41 $\sin$ 0.99 $\lambda$ & $\delta$ = 7.06 - 3.37 $\cos$ $\lambda$ + 0.15 $\sin$ $\lambda$ \\
Ganymede & $\delta$ = 12.45 + 6.16 $\cos$ 0.98 $\lambda$ + 3.64 $\sin$ 0.98 $\lambda$ & $\delta$ = 12.97 - 5.39 $\cos$ 0.99 $\lambda$ - 4.50 $\sin$ 0.99 $\lambda$ \\
\hline 
\end{tabular}
\caption{Best fit of the northern and southern lead angles for Io, Europa and Ganymede, with $\lambda$ the SIII West-longitude of the moon.}
\label{tab:fit_param}
\end{center}
\end{table}

\begin{figure}[h!]
\centering
\includegraphics[width=0.9\columnwidth]{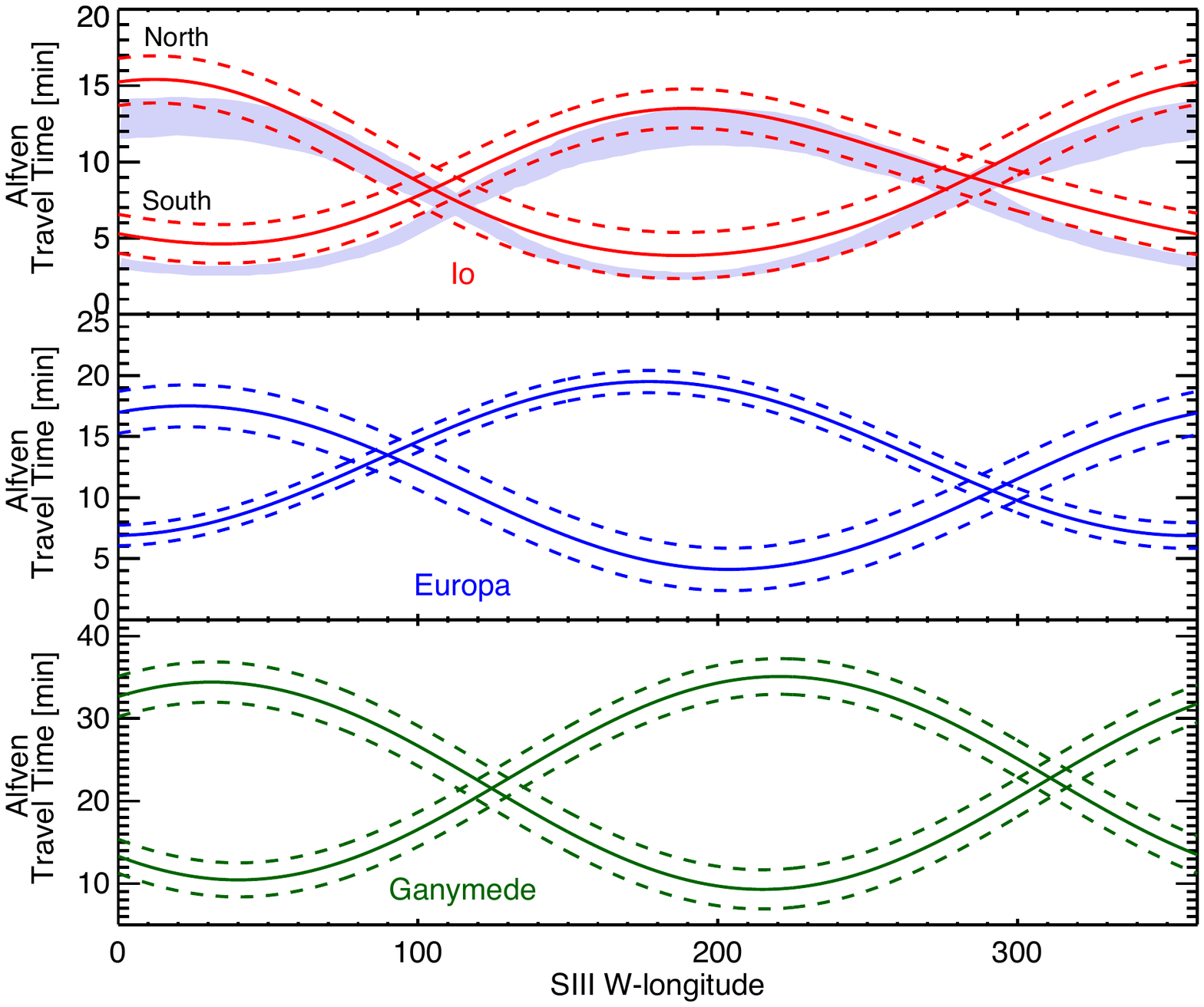}
\caption{Alfv\'en travel times for Io (top), Europa (middle) and Ganymede (bottom), estimated from the equatorial lead angles measured by UVS. Blue-shaded area for Io comes from \citeA{Hinton2019}.}
\label{fig:ATT}
\end{figure}

The Alfv\'en travel times can be approximated using measurements of the equatorial lead angle. The quantity $\delta_{moon}$/360 is the orbital fraction over which the Alfv\'en wave travels from the interaction region to the Jovian ionosphere. If the moons were static, or far slower than the angular rotation of Jupiter, the Alfv\'en travel times could be estimated by multiplying this ratio by the Jovian orbital period. However, since the orbital period of the moon is not negligible when compared with the Jovian rotation period (especially for Io), the synodic period of the moon has to be accounted for. The Alfv\'en travel times can then be approximated using equation (\ref{eq:Att}).

\begin{equation}
 t_{A} = \frac{P^{syn}_{moon} \times \delta_{moon}}{360},
 \label{eq:Att}
\end{equation}

where $P^{syn}_{moon}$ is the synodic period and $\delta_{moon}$ the measured equatorial lead angle of the moon of interest. The range of Alfv\'en travel times are listed in Table \ref{tab:Alven_TT}, and displayed on Figure \ref{fig:ATT}. The derived Io Alfv\'en travel times are consistent with the modeled travel times by \citeA{Hinton2019}.

\begin{table}[ht]
\begin{center}       
\begin{tabular}{c|c|c} 
 & North & South  \\
 \hline
Io & $t_{A}^{min}$ = 3.9 $\pm$ 1.5\,minutes & $t_{A}^{min}$ = 4.6 $\pm$ 1.3\,minutes   \\
   & $t_{A}^{max}$ = 15.4 $\pm$ 1.5\,minutes & $t_{A}^{max}$ = 13.5 $\pm$ 1.3\,minutes \\
 \hline
Europa & $t_{A}^{min}$ = 4.1 $\pm$ 1.7\,minutes & $t_{A}^{min}$ = 6.9 $\pm$ 1.0\,minutes    \\
       & $t_{A}^{max}$ = 17.5 $\pm$ 1.7\,minutes & $t_{A}^{max}$ = 19.5 $\pm$ 0.9\,minutes  \\
 \hline
Ganymede & $t_{A}^{min}$ = 9.3 $\pm$ 2.4\,minutes & $t_{A}^{min}$ = 10.4 $\pm$ 2.0\,minutes  \\
         & $t_{A}^{max}$ = 34.4 $\pm$ 2.4\,minutes & $t_{A}^{max}$ = 35.1 $\pm$ 2.2\,minutes \\
\hline 
\end{tabular}
\caption{Range of northern and southern hemisphere Alfv\'en travel times for Io, Europa and Ganymede estimated from the measured lead angles.}
\label{tab:Alven_TT}
\end{center}
\end{table}

The lead angle shows considerable variability at a given SIII longitude, well beyond the observational uncertainties. Since this quantity accounts for the sum of the physical processes occurring between the interaction region around the moon and Jupiter's ionosphere, it is affected by plasma density variations encountered by the Alfv\'en waves within the torus and potential magnetic field strength variations. For Io, \citeA{Hinton2019} estimated that the Alfv\'en travel times to Jupiter's southern ionosphere at $\lambda^{Io}$ = 200$^{\circ}$ ranged from 11\,min to 13.5\,min solely based on torus density variations on the order of 50\% caused by variation in Io's volcanic activity \cite{Yoshikawa2017}. Based on Io's synodic period, a 2.5\,min variation in Alfv\'en travel times corresponds to a longitudinal shift of about 1.5$^{\circ}$, which is fully consistent with the lead angle variations observed between perijoves. However, \citeA{Hinton2019} also predicted a variation in the Alfv\'en travel times to Jupiter's southern ionosphere at $\lambda^{Io}$ = 30$^{\circ}$ from 2\,min to 3\,min, leading to a predicted shift in lead angle of 0.6$^{\circ}$, which appears smaller than the measured variation in lead angle in that longitude sector. The origin of this discrepancy is not clear, and could be due to either underestimating the uncertainties of the lead angle, uncertainties in the magnetic field model or simplifications in modeling the plasma sheet spatial variability.

\section{Variability in magnetospheric conditions and lead angle}
\label{sec:variability}

The scatter in lead angle shown in figures \ref{fig:Northern_LA} and \ref{fig:Southern_LA} implies significant variations in the magnetospheric conditions at the moons. Because the lead angle depends on the Alfv\'en travel times from the interaction region to the ionosphere, any temporal or spatial changes in the plasma density, structure of the plasma sheet, or even magnitude of the magnetic field in these regions need to be considered. We briefly discuss here the spatial and/or temporal variations of such quantities.

Neutrals escaping Io are dissociated and ionized by impacts from magnetospheric electrons \cite{Bagenal2020}. They are then picked up by the rotating magnetic field and brought up to near-corotation by transferring angular momentum from the Jovian ionosphere to the magnetosphere through a system of field-aligned currents, and regulated by the ionospheric conductivity \cite{Hill1979}. The radial transport of plasma then occurs through the centrifugally-driven flux tube interchange, in which plasma-loaded flux tube moves radially outward from its production source by the \textbf{E} $\times$ \textbf{B} drift, and are replaced by inward-moving flux tubes containing less mass \cite <e.g.,>[]{Thomas2004}. The typical transport timescale of plasma from the Io to the Europa orbital distances was estimated to be in the $\sim$30-80 days \cite{Bagenal2011}.

Plasma survey at various distances across the magnetosphere have been performed thanks to the previous Jupiter missions, and is being extended now with Juno \cite <see, e.g., reviews of>[]{Bagenal2015, Bagenal2020}. Density in the plasma sheet generally decreases by five orders of magnitude from 6\,R$_J$ to 30\,R$_J$ \cite{Bagenal2016}. Because of Jupiter's magnetic dipole tilt, magnetic field measurements from in-situ spacecrafts are characterized by regular sign changes in the radial component of that field associated with a plasma sheet crossing. Although previous in-situ plasma measurements performed in the middle magnetosphere of Jupiter showed a well-structured plasma sheet, variations in magnetic field measurements have been reported, indicating spatial and temporal variations of the current sheet position \cite <e.g.,>[]{Krupp2004}.

Since the early Pioneer 10 and 11 missions, the magnetic field has been known to be distorted near the centrifugal equator, where the field is radially stretched outwards by the presence of the plasma sheet currents. An empirical model of the magnetodisk using the Voyagers measurements that accounts for a system of azimuthal currents circulating in the centrifugal equator were constructed to account for this effect \cite{Connerney1981}. The extensive data collected by Juno demonstrates that the magnetic field in the middle magnetosphere is less stretched than previously thought. This led \citeA{Connerney2020} to revise the empirical magnetodisk model, and to add a radial current that contributes to the azimuthal component of the magnetic field (B$_{\Phi}$), and designed to account for the transfer of angular momentum of the radially outflowing plasma. Orbit-by-orbit measurements of the radial current system contributing to the B$_{\Phi}$ component show significant variations, which might suggest a modulation in the angular momentum transfer that affects the plasma outflow \cite{Connerney2020}. \citeA{Vogt2022b} used magnetic field measurements from Juno as well as previous missions to survey the magnetic field condition near the orbit of Ganymede. They found that the expected temporal variability obtained from fit of the current sheet results in a 10-20\,\% variability in the magnetic field components, and is longitude-dependent. Understanding and tracking the variability of the plasma sheet current is crucial as they may induce a displacement of the footprint location, especially for Ganymede \cite{Vogt2022a, Promfu2022}. The variation in the Ganymede footprint position are large enough to be easily detected and thus is important to track magnetospheric changes.

\citeA{Huscher2021} used Juno-JADE data at or near the plasma sheet crossing to survey the plasma density from Juno's first 26 orbits at radial distances beyond 17\,R$_J$. The plasma sheet density structure can show significant variability in spatial structure, sometimes smaller than 1\,R$_J$, over timescale of minutes, and between Juno orbits. At the same radial distance within the plasma sheet, the peak charged density of heavy ions measured by Juno-JADE can differ by about one order of magnitude between orbits. For instance, data from a few number of orbits indicated plasma density uniformly low (\textit{e.g.,} PJ12 and PJ26). Juno will be traversing the equatorial Io-Europa region as the extended mission progresses, and more results are expected.

\begin{figure}[h!]
\centering
\includegraphics[width=1.02\columnwidth]{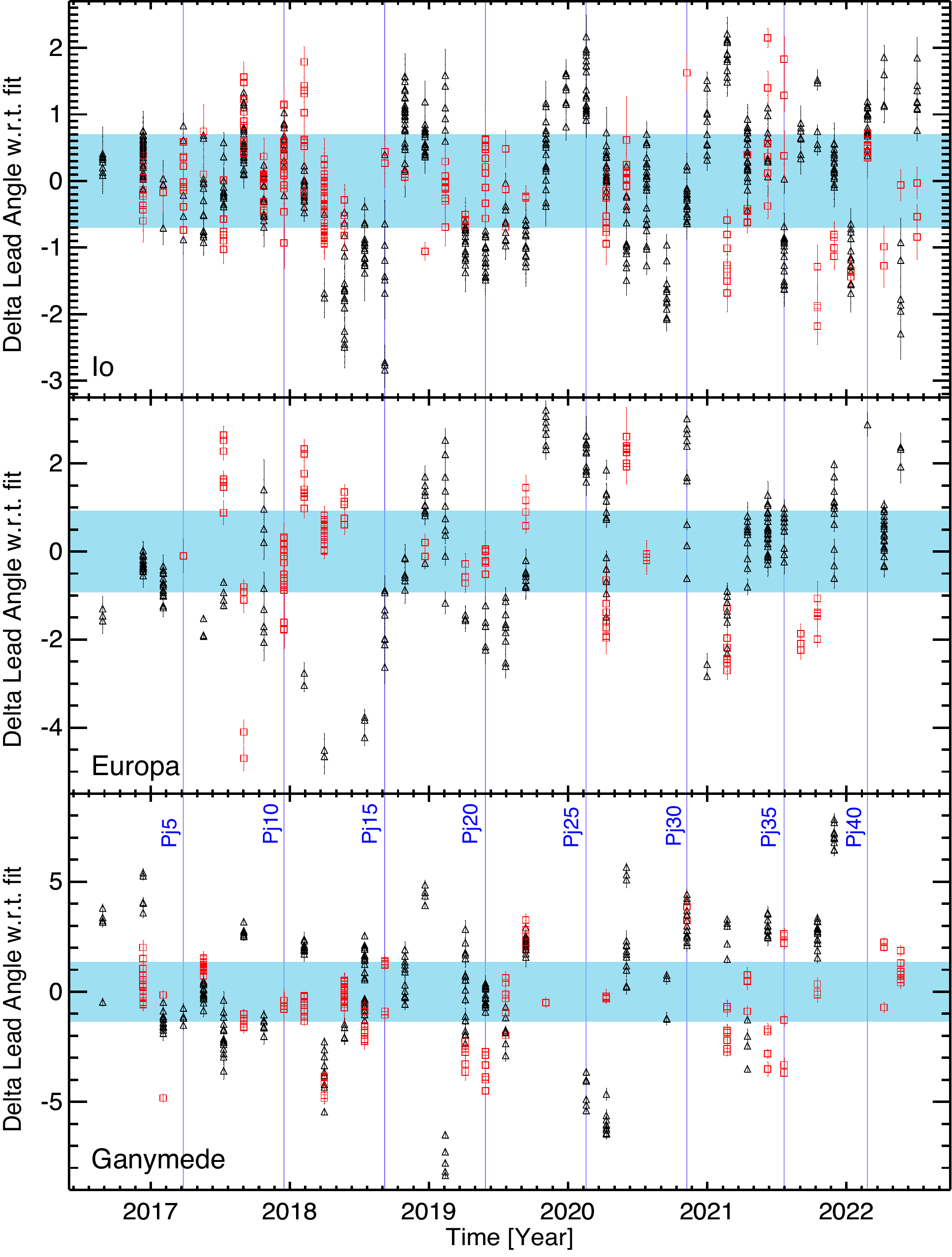}
\caption{Time series of the measured lead angle deviation from the best fit shown in figures \ref{fig:Northern_LA} and \ref{fig:Southern_LA}. Red squares and black triangles correspond to the northern and southern lead angle data. The blue-shaded regions correspond to the uncertainty from the northern lead angle fit parameters shown in figure \ref{fig:Northern_LA}.}
\label{fig:Delta_LA}
\end{figure}

In order to investigate temporal variability in the plasma sheet properties, figure \ref{fig:Delta_LA} present the lead angle deviations from the best fit obtained on figures \ref{fig:Northern_LA} and \ref{fig:Southern_LA}. Juno-UVS lead angle measurement is subtracted from the lead angle fit at the particular $\lambda$ the data was recorded at. A positive/negative value for that quantity, called here $\Delta_{LA}$, corresponds to the case where the Alfv\'en travel time was longer/shorter than the travel time derived from equation (\ref{eq:Att}), respectively. A longer Alfv\'en travel time may translate into the situation where either (i) the local plasma sheet density was higher than nominal, (ii) the plasma sheet vertical extent increased, (iii) the local magnetic field amplitude decrease with respect to it nominal value, or a combination thereof.

The consistently lower density reported by \citeA{Huscher2021} using Juno-JADE during the inbound section of Juno's orbit 12 throughout the magnetosphere may cause the Alfv\'en travel time to become temporarily shorter for that orbit, until the plasma density return to their nominal level. This would translate to a negative $\Delta_{LA}$, and is roughly consistent with the $\Delta_{LA}$ measured for Io (south), Europa (south) and Ganymede (both north and south) on PJ12. A couple of issues with that interpretation are that (i) Juno-JADE only measured the densities from 17-50\,R$_J$, \textit{i.e.,} from Ganymede and beyond, (ii) the JADE data were recorded several days prior to the UVS data (1-3 days, depending on the orbit number). Lower plasma density in the middle magnetosphere measured by JADE may be a consequence of a temporary slowdown of the mass loading from Io, implying lower densities in the inner magnetosphere prior to PJ12. This means that UVS should have recorded consistently negative $\Delta_{LA}$ on, \textit{e.g.,} PJ11, which is not the case.

Although no clear temporal trends can be distinguished from figure \ref{fig:Delta_LA}, a Lomb–Scargle periodogram analysis on these time-series reveals a period in the 400-500\,days range for Io. This may be consistent with the periodic brightening of Loki Patera, the most powerful volcano on Io, as monitored in the infrared. The periodicity of such brightness was thought to be correlated with periodic changes in Io's eccentricity and semimajor axis, about 480 and 460 days, respectively \cite{DeKleer2019}. Io is the main supplier of plasma in the Jovian magnetosphere. Changes in Io's volcanic activity affect the radial circulation of mass and energy in the middle magnetosphere over time scales of tens of days. For instance, \citeA{DeKleer2016} have characterized the location and appearance of hot spots on Io in the near-IR and found widespread activity from Aug–Sep 2013 and from Oct 2014-May 2015. Around the same time, the Hisaki telescope monitored the torus ion emission lines of the Jovian system in the 55–145\,nm range \cite{Yoshikawa2014}. Enhancements in the sodium emission line  \cite{Yoneda2015}, as well as sulfur and oxygen emission lines \cite{Tsuchiya2018} was detected from mid-Jan until mid-Mar 2015, over sensibly different timescales. During the Cassini era, temporal changes in the emissions from  the major sulfur and oxygen ions torus species was suggested to be related to the changes in outgassing from Io \cite{Delamere2004, Steffl2006}. However, tying together changes in the brightness of Io's volcanoes with enhancements in the various emission lines of the torus (\textit{e.g.,} sodium, oxygen and sulfur), to ultimately derive changes in mass supplied to the torus is not straightforward and still unclear.

The present work provides an empirical fit of the equatorial lead angle for Io, Europa and Ganymede derived from Juno data, from which the averaged Alfv\'en travel time is derived. It also demonstrates how, by accounting for these lead angle values, the interpretation of the moon-induced decametric radio emissions can be improved, as demonstrated below. Understanding the effect of magnetospheric conditions on the lead angle variability is beyond the scope of this work, and deserves a dedicated study.

\section{Application to modeling of the moon-induced decametric emission}
\label{sec:decametric}

The knowledge of the lead angle is also important for the interpretation of the moon-induced decametric emissions \cite <e.g.,>[]{Hess2010b, Marques2017, Louis2019, Lamy2022}. This section illustrates the added benefit from the equatorial lead angle knowledge in a case related to the interpretion of the Ganymede-induced decametric radio emission. These emissions are thought to originate from the cyclotron maser instability (CMI), a wave-particle instability where a circularly polarized wave resonates with the gyrating motion of an accelerated electron population \cite{Wu1979, Wu1985, Treumann2006}. That mechanism was confirmed using Juno in-situ measurements \cite{Louarn2017, Louarn2018, Louis2020}. These radio emissions are produced along the magnetic field lines, at a frequency close to the electron cyclotron frequency (proportional to the magnetic field amplitude), and beamed along the edges of a thin hollow cone on the order of a degree in thickness. The opening angle of this cone is dependent on the electron energy \cite{Hess2008}. In order to interpret the moon-induced decametric emission, modeling tools such as the Exoplanetary and Planetary Radio Emission Simulator \cite <ExPRES, >[]{Hess2008, Louis2019} require knowledge of both the lead angle and electrons velocity. Because the directionality of these radio emissions depend both on the electron energy and the lead angle, this may lead to situations where non-unique solutions exist, in order to reproduce the observed decametric radio arcs. The knowledge of the active field line (where the radio emission sources are located) and thus the lead angle, constitutes one of the main sources of uncertainty in determining the electron energy responsible for the radio emission \cite{Hess2010b, Lamy2022}. Previous estimate of the absolute lead angle values were not accurate enough mostly because of the uncertainty in the magnetic field models as well as the uncertainties in the Europa and Ganymede measured footprint positions from HST \cite{Hess2010b}, leading to non-physical cases with negative lead-angle.

\begin{figure}[h!]
\centering
\includegraphics[width=1.15\columnwidth]{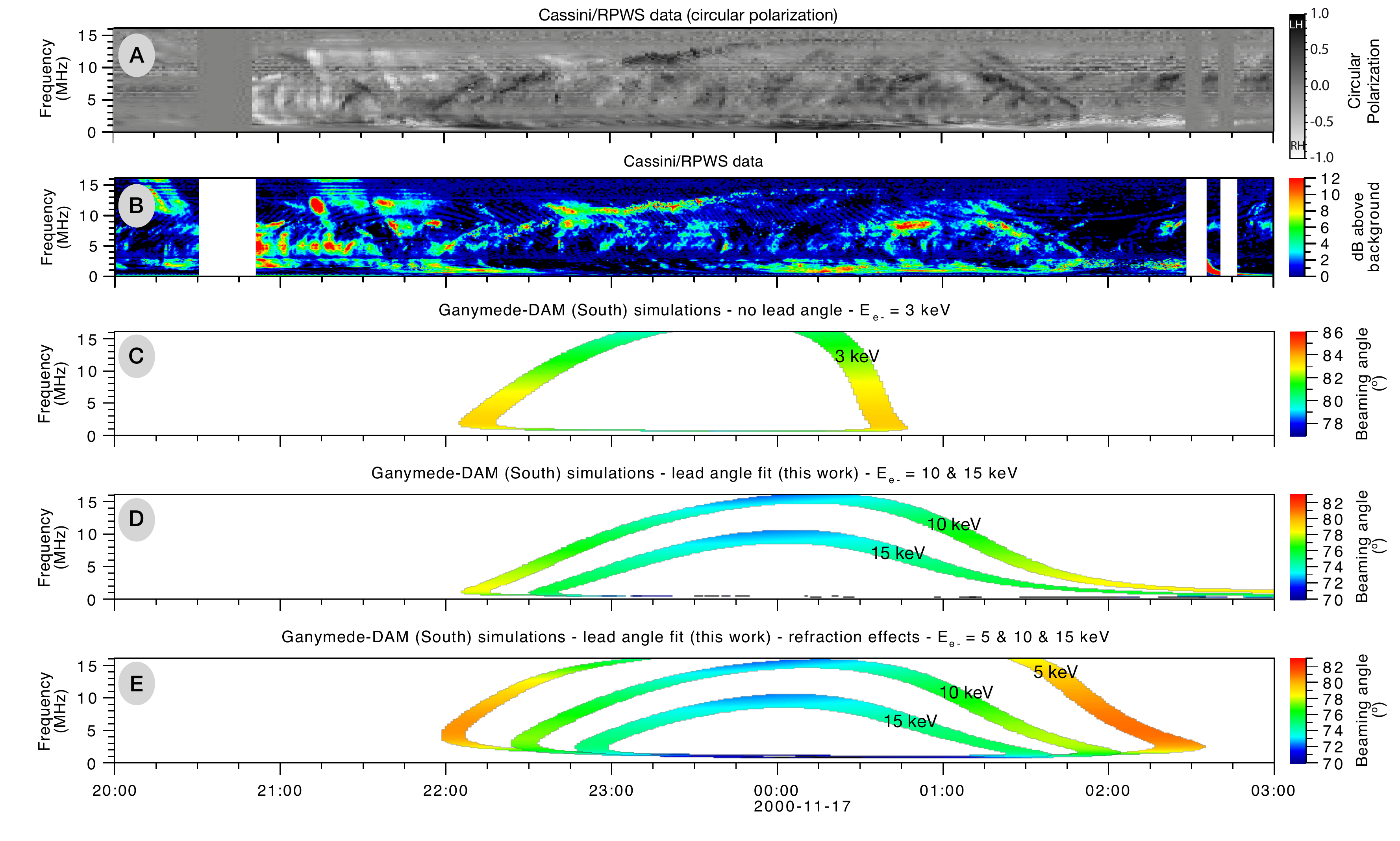}
\caption{Panel A: Cassini-RPWS time-frequency spectrogram of circularly polarized radio emission. Panel B: Cassini-RPWS time-frequency spectrogram. Panel C: ExPRES simulation of a Ganymede-D decametric arc without lead angle. Panel D: ExPRES simulation using the lead angle derived from this present study. Panel E: ExPRES simulation combining the lead angle derived from this present and refraction effect on the emission. This simulations have been produced using the Version 1.2.0 of ExPRES \cite{louis_corentin_k_2023_7759511}}.
\label{fig:Decametric_Ganymede}
\end{figure}

Several datasets of moon-induced decametric radio emission were recorded by the Radio and Plasma Wave Science (RPWS) instrument on Cassini \cite{Gurnett2004} on 17 November 2000, prior to the Jupiter flyby. Figure \ref{fig:Decametric_Ganymede} shows the analysis of a decametric arc induced by Ganymede, and previously studied by \citeA{Louis2017}. RPWS data is shown on panels A and B as time-frequency spectrograms, corresponding to a Ganymede-D decametric arc, \textit{i.e.,} emitted on a field line connecting Ganymede and Jupiter's southern hemisphere around Jupiter's dawn side. Because the waves amplified by the cyclotron maser instability are circularly polarized, panel A shows the time-frequency spectrogram corresponding to the degree of circular polarization, for better contrast. During the observation of the Ganymede-D arc (from $\sim$22:00 to $\sim$02:00), the longitude of Ganymede varies from 348$^{\circ}$ to 133$^{\circ}$.

Panel C shows the best fit to the data, prior to this work, using the ExPRES radio modeling \cite{Louis2019}. This assumes (i) the radio emission is created by a loss-cone electron distribution with energies of 3\,keV, (ii) the JRM33 magnetic field model, and (iii) no equatorial lead angle. Note that we consider here a simplistic mono-energetic case, as Juno clearly observed electrons distributed as a power-law extending well above and below 3\,keV \cite <e.g.,>[]{Szalay2018}. Although it was previously presented by \citeA{Louis2017}, this application differs slightly from that earlier work which assumed the ISaAC magnetic field model \cite{Hess2017}. Panel D shows simulations using ExPRES considering the equatorial lead angle fit derived from the present work. The best fit of the data is obtained considering electrons with energies ranging from 10-15\,keV. When accounting for refraction effects on the beamed radio emissions, the fit can be improved by considering electrons of energies ranging 5 to 15\,keV (panel E). Refraction effects are occuring when the radio emissions are emitted in a region with decreasing electron cyclotron frequency (f$_{ce}$). In which case, the emission can not propagate due to the cut-off frequency at f$_{ce}$, and the radio wave is reflected until it reaches a region with increasing f$_{ce}$ \cite{Galopeau2016, Louis2017b, Louis2021}.

The addition of the knowledge on the equatorial angle, especially for Europa and Ganymede, improves the predictability and analysis of moon-induced decametric radio emission. Due to Ganymede's longitude variation during the observation of the Ganymede-D arc, based on our new lead angle model, the error on the active magnetic field line position goes from 5$^{\circ}$ (minimal value of $\delta$ in the south) to 14$^{\circ}$, which lead to a poor estimate on the location of the sources. Since the shape of the arc depends mainly on the lead angle (position of the active field line), the magnetic field model (value of B, and therefore position of the sources, along the field line) and the energy of the electrons (value of the emission cone aperture), a too low estimate of the longitude of the active field line will lead to an underestimation of the value of the energy of the electrons \cite{Hess2010b, Louis2017c}. This is what is noted in the simulations comparing figures \ref{fig:Decametric_Ganymede}A and \ref{fig:Decametric_Ganymede}C, where a lower value of electrons is needed to fit correctly the arc without a lead angle model.

This improved estimation of the position of the active field lines, and therefore of the position of the radio sources, allows reducing the number of unknowns and to obtain a better, and more reliable, estimate of the electron energy. For instance, comparing in more details the simulations that account for the lead angle and refractions effects (figure \ref{fig:Decametric_Ganymede}C) with the circularly polarized radio emission from RPWS indicates that the early part of the radio emission arc (from 22:00:00 to 22:15:00) are better reproduced with electrons with lower energies, \textit{i.e.,} 5\,keV, while the center and end of the arc are better reproduced with electrons with energies ranging from 10 to 15\,keV. It is worth noting here that \citeA{Lamy2022} showed that, in the case of Io, the electrons energy derived from radio emission analysis is variable as a function of the moon longitude. It is not surprising here to observe that in order to reproduce the arc in the time-frequency plane, the energy of the electrons must vary with time (\textit{i.e.,} the longitude of the moon). Finally, the ExPRES simulations with the new lead angle model give results that are in agreement with the Juno in-situ observations recorded during a Ganymede fluxtube crossing on PJ20, on May 29$^{th}$ 2019, \cite{Louis2020}. 

\section{Discussion and Summary}

The presented work makes use of 479 and 1148 individual spectral images of the satellite footprints recorded over the northern and southern auroral regions, respectively, measured by Juno-UVS from PJ1 to PJ43 (at the exception of PJ2), \textit{i.e.,} about 6 years of Juno data. From these images, the accurate positions of the Io, Europa and Ganymede Main Alfv\'en Wing spots are estimated and compared with the prediction from the magnetic field model obtained at the end of Juno's prime mission, JRM33 \cite{Connerney2022}. The accuracy with predictions from JRM33 can be estimated while limiting the projection effect by restricting to data recorded at small emission angles. We selected data points with emission angle lower than 20$^{\circ}$, which allowed us to reduce the uncertainty associated to the projected vertical extent of the footprint emission curtain ($\sigma_P$ in \ref{eq:2}) while keeping enough data to perform a statistical analysis. In general, the average distance between the JRM33-computed moon footpaths and the observed MAW positions is smaller than 500\,km, which is equivalent to $<$\,0.4$^{\circ}$ on Jupiter.

Measurements of the equatorial lead angle provide information on how the Alfv\'en waves generated around the moon interaction region propagate towards the Jovian ionosphere. It is an important piece of information for (i) interpreting the footprint related in-situ measurements made by Juno, (ii) inferring in-situ conditions of the plasma sheet (\textit{e.g.}, a lead angle greater than the fit derived here might be caused by a higher than usual plasma density, or a variation in the magnetic field strength), and (iii) analyzing the satellite-induced decametric radio emission. Despite the variability in the measured lead angle at a given moon longitude, a statistically consistent trend can be derived from the first 43 perijoves. The best fit of the lead angle for Io is in the 1.8$^{\circ}$-7.2$^{\circ}$ range, with an average calculated as the arithmetic mean of 4.2$^{\circ}$. For Europa, the lead angle ranges from 2.2$^{\circ}$ to 10.4$^{\circ}$, with an average of 6.4$^{\circ}$. For Ganymede, the lead angle is in the 5.3$^{\circ}$-20.0$^{\circ}$ range, with an average of 12.7$^{\circ}$. Over an entire Jovian rotation, this corresponds to Alfv\'en travel times ranging from 3.9-15.4\,minutes, 4.1-19.5\,minutes, and 9.3-35.1\,minutes for Io, Europa and Ganymede, respectively. 

Significant deviations of the lead angle at a given longitude were observed for all moons. Changes in the magnetospheric conditions, such as plasma density variations, magnetic field amplitude or spatial structure of the plasma sheet, are likely responsible for such variations. Comparing the instantaneous lead angle obtained from the observed MAW footprint positions with the fit derived may provide an indication of these changes in magnetospheric conditions. Time series of the lead angle variation with respect to the best fit does not show a consistent temporal trend for Europa and Ganymede (see figure \ref{fig:Delta_LA}). For Io, a Fourier analysis shows a tentative periodic change in the 400-500\,days range, which may be attributed to modulated in Io's volcanic activity, although a dedicated study on this topic is beyond the scope of this investigation and deserves a detailed examination.

Modeling the moon-induced decametric arc requires the knowledge of the equatorial lead angle to constrain with higher accuracy the derived electron energy causing the emission. When accounting for the equatorial lead angle values from this work in the ExPRES simulation tool, the derived electron energy required to reproduce the Ganymede-induced decametric arc is in agreement with the in-situ measured values from Juno. This method can be applied for modeling the moon-induced radio emission for which no in-situ particle measurements data is available.

The main conclusions of this work are listed as follows:
\begin{enumerate}
    \item The reported position of the Main Alfv\'en wing spots for Io, Europa and Ganymede agrees well with the JRM33-computed satellite footpath generally $<$\,500\,km, which is equivalent to $<$\,0.4$^{\circ}$ on Jupiter.
    \item Empirical formulae for the Io, Europa and Ganymede equatorial lead angles derived from Juno data are provided.
    \item The range of equatorial lead angle for Io, Europa and Ganymede is 1.8$^{\circ}$-7.2$^{\circ}$, 2.2$^{\circ}$-10.4$^{\circ}$, 5.3$^{\circ}$-20.0$^{\circ}$, respectively.
    \item The respective range of Alfv\'en travel times for Io, Europa and Ganymede, derived from the lead angle, are 3.9-15.4\,minutes, 4.1-19.5\,minutes, and 9.3-35.1\,minutes.
    \item Knowledge of the lead angle allows, \textit{e.g.,} better simulating the decametric radio emission induced by the Galilean moons. By comparing them with the observations, this makes possible deriving more precisely the energies of the electrons triggering these emissions. This new derivation is in agreement with the Juno in-situ measurements.
    
\end{enumerate}

\section{Open Research}

All the data used in this study are publicly available on the PDS Atmospheres Node Data Set Catalog \url{https://pds-atmospheres.nmsu.edu/cgi-bin/getdir.pl?dir=DATA&volume=jnouvs_3001}. Juno-UVS calibrated data (Reduced Data Record) recorded during Juno's perijove observation sequences were used here. The corresponding dataset used here contain the string \textit{PXXOBS}, where XX = 1 to 43. Cassini/RPWS data display in this article are part of the \citeA{Cecconi2017} collection, and has been produced following the "Circular Polarization mode" goniopolarimetric Inversion described in section 2.1.3.2 of \citeA{Cecconi2005}.

\acknowledgments

We are grateful to NASA and contributing institutions that have made the Juno mission possible. This work was funded by the NASA's New Frontiers Program for Juno via contract NNM06AA75C with the Southwest Research Institute. CKL's work at the Dublin Institute for Advanced Studies was funded by Science Foundation Ireland Grant 18/FRL/6199. AHS acknowledges NASA NFDAP grant 80NSSC23K0276. Hue acknowledges support from the French government under the France 2030 investment plan, as part of the Initiative d’Excellence d’Aix-Marseille Université – A*MIDEX AMX-22-CPJ-04.


\end{document}